\documentclass[aip,reprint]{revtex4-1}
	\usepackage{amsmath}
	\usepackage{makeidx}
	\usepackage{mathtools}
	\usepackage{tabularx}
	\newcolumntype{Y}{>{\centering\arraybackslash}X}
	\usepackage{booktabs}
	\usepackage{hhline}
	\usepackage{amsfonts}
	\usepackage[ansinew]{inputenc}
	\usepackage[usenames,dvipsnames]{pstricks}
	\usepackage{subfigure}
	\usepackage{multirow}
	\usepackage{array}
	\usepackage{epsfig}
	\usepackage{pst-grad} 
	\usepackage{pst-plot} 

\makeindex

\begin{document}

\title{Fast relaxation of photo-excited carriers in 2D transition metal dichalcogenides}

\author{Mark Danovich}
\email{mark.danovich@postgrad.manchester.ac.uk}
\affiliation{National Graphene Institute, University of Manchester, Booth St E, Manchester M13 9PL, UK}
\author{Igor Aleiner}
\affiliation{Physics Department, Columbia University, New York, NY 10027, USA}
\author{Neil D. Drummond}
\affiliation{Department of Physics, Lancaster University, Lancaster LA1 4YB, United Kingdom}
\author{Vladimir Fal'ko}
\affiliation{National Graphene Institute, University of Manchester, Booth St E, Manchester M13 9PL, UK}

\date{\today}

\begin{abstract}
We predict a fast relaxation of photo-excited carriers in monolayer transition metal dichalcogenides (TMDCs), which is mediated by the emission of longitudinal optical (LO) phonons. By evaluating Born effective charges for ${\rm MoS_2, MoSe_2, WS_2,}$ and ${\rm WSe_2}$, we find that, due to the polar coupling of electrons with LO phonons, the relaxation times for hot electrons and holes are of the order of a few ${\rm ps}$.
\end{abstract}


\maketitle
\begin{figure}[!t]
\centering
\includegraphics[width=0.35\textwidth]
{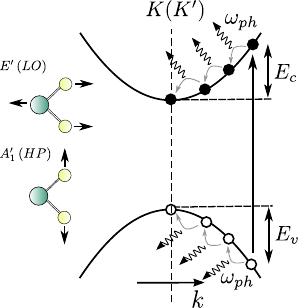}
\caption{Sketch of the energy relaxation of photo-excited carriers in the valence (v) and conduction (c) bands of TMDCs through phonon emission. The use of the parabolic approximation in the description of electron and hole dispersion in each valley ($K$ or $K'$) sets a constraint, $E\le 0.7\,{\rm eV}$ on the excitation energies of the charge carriers. The insets show side view of the atomic displacements in the LO and HP modes.}
\label{fig:phonon_scat}
\end{figure}

Monolayer transition metal dichalcogenides (TMDCs) 
offer a unique possibility to create nm-thin optoelectronic devices\cite{devices,tmdcs_opto, Heinz, xu,mak,xiao,sallen,cao,zeng}, in particular when used in van der Waals heterostructures with other two-dimensional (2D) crystals \cite{geim_hetero}. The optoelectronic functionality of TMDCs is determined by their high-efficiency optical absorption in the visible optical range \cite{kdotp} as well as the fact that their monolayers are direct-band-gap 2D materials. Because of their promise for optoelectronics, it is important to understand the process of cooling (energy relaxation) of photo-excited carriers in TMDCs. In this paper we show that photo-excited carriers relax inelastically on a ${\rm ps}$ time scale by emitting $\Gamma$-point optical phonons. Such a high speed of relaxation of electrons and holes excited to energies $>100\,{\rm meV}$ above the band edge arises from polar coupling to the longitudinal optical (LO) phonons in the 2D crystal.
In the theory reported in this Letter, we analyse the phonon-mediated cooling of hot electrons/holes in TMDCs, taking into account two phonon modes coupled to the intra-band intra-valley relaxation processes: the in-plane LO phonon and the out of plane ``homopolar'' (HP) vibrational mode. These are the only vibrational modes coupled to the electron (hole) intravalley transitions, whereas density functional theory (DFT) modelling produces electron (hole) couplings to the corner of the Brillouin zone (K-point phonons), which are weaker by at least an order of magnitude\cite{defo_params,defo_params2,X,Y}.
The electron-phonon interaction in TMDCs is given by the Hamiltonian
\begin{equation}
H_{e-ph}=\sum_{\mathclap{\substack{\vec{q},\vec{k}\\ \mu=LO,HP}}}
 g_{\mu,\vec{q}}c^{\dagger}_{\vec{k}+\vec{q}}c_{\vec{k}}(a^{\dagger}_{\mu, -\vec{q}}+a_{\mu, \vec{q}}),
\label{eq:hamil}
\end{equation}
where $c_{\vec{k}}^{\dagger}\,(c_{\vec{k}})$ are the creation (annihilation) operators for a charge carrier (electron or hole) in the vicinity of one of the valleys, $K$ or $K'$, near the corners of the hexagonal Brillouin zone of the 2D crystal \cite{yao_review, optical_rev}, with $\vec{k}$ measured from the valley center (see Fig.\,\ref{fig:phonon_scat}). The operators $a_{\mu,\vec{q}}^{\dagger}\,(a_{\mu,\vec{q}})$ are the phonon creation (annihilation) operators for mode $\mu={\rm LO}$  or ${\rm HP}$ with wavevector $\vec{q}$, where $|q|\ll|K|$. 
The two phonon modes \footnote{TMDCs have 6 optical modes denoted by the irreducible representations of the point group $D_{3h}$ ($A_1', A_2'', E', E''$), and 3 acoustical modes denoted by LA, TA, and ZA, where LA and TA are the in-plane longitudinal and transverse modes and ZA is the out-of-plane mode. We neglect the transverse optical and acoustical modes due to their weak coupling at the $\Gamma$ point, $q\rightarrow 0$} accounted for in the relaxation process are shown in Fig.\,\ref{fig:phonon_scat}. 

The LO mode, which corresponds to the irreducible representation $E'$ of the symmetry group $D_{3h}$ of the crystal, couples to the charge carriers through the polarization\footnote{We note that the polarization induced by the LO phonon mode can also interact with itself, leading to a correction to the optical phonon spectrum. Near the $\Gamma$ point, for $qr_*\ll 1$, the dispersion of 2D LO phonons can be approximated by $\omega=\omega_{LO}+aq$, where $a$ is a constant. The correction to the optical phonon spectrum from the polarization interaction comes from the term in the energy density in reciprocal space given by $\frac{1}{2}\int \frac{d^2q}{(2\pi)^2} \frac{|\vec{q}\cdot \vec{P}(\vec{q})|^2}{q(1+qr_*)}$. This then gives a correction to the constant phonon frequency near the $\Gamma$ point, $\omega=(\omega_{LO}^2+\frac{Z^2e^2}{M_{r}A}\frac{q}{1+qr_*})^{1/2}$. In our cases, $\frac{Z^2e^2}{\omega_{LO}^2 M_r A}\frac{q}{1+qr_*}\ll 1$ such that for $qr_*\lesssim 1$ we can write $\omega=\omega_{LO}+aq$, with $a=\frac{Z^2e^2}{2\omega_{LO} M_r A}$, but due to $a/r_*\ll \omega_{LO}$, this correction to the phonon dispersion can be neglected.} induced by the lattice deformation, $\vec{P}=\frac{Z e}{A}\vec{u}$, where $Z$ is the Born charge, $\vec{u}$ is the relative displacement of the two sublattices in the LO vibration, $A$ is the unit cell area, and $e$ is the electron charge, leading to
\begin{equation}
g_{LO}=\frac{i}{A}\sqrt{\frac{\hbar}{2NM_r\omega_{LO}}}\frac{2\pi Z e^2}{1+qr_*},
\label{eq:glo}
\end{equation}
where $N$ is the number of unit cells, $M_r$ is the reduced mass of the two sublattices, and $\omega_{LO}$ is the LO phonon frequency. The dielectric screening of the electric field of LO mode deformations is described \cite{ganchev,Heinz} by the factor $1/(1+qr_*)$, where $r_*$ is a length scale defined by $r_*=a_z(\epsilon_{||}-1)/2$, where 
$a_z$ and $\epsilon_{||}$ are the $z$-axis lattice constant and in-plane dielectric constant of a bulk crystal of the corresponding TMDC\cite{ganchev}. In the absence of established values of $\epsilon_{||}$ for undoped bulk crystals of TMDCs, below we shall treat $r_*$ as a phenomenological parameter, with values\cite{Heinz} in the range $r_{*} \sim 15\--30$.
\begin{table}[!t]
\centering
\caption{TMDC parameters used in the modelling of phonon emission rates: the carrier masses expressed in terms of the free electron mass $\frac{m_e}{m_0}$ and $\frac{m_v}{m_0}$, the unit cell area $A$, the reduced mass in the unit cell $\frac{M_r}{m_p}$ expressed in terms of the proton mass, the total atomic mass in the unit cell $\frac{M}{m_p}$, the optical phonon frequencies $\hbar\omega_{LO}, \hbar\omega_{HP}$, and the deformation potentials $D_{c(v)}$ for electrons in the conduction (c) and valence (v) bands.}
\vspace{0.2cm}
\label{params}
\begin{tabular}{lcccc}
\hline\hline
& ${\rm MoS_2}$        & ${\rm MoSe_2}$       & ${\rm WS_2}$         & ${\rm WSe_2}$        \\ \hline
$\frac{m_e}{m_0}$ [Ref. \onlinecite{kdotp}] & $0.46$               & $0.56$               & $0.26$               & $0.28$               \\
$\frac{m_v}{m_0}$ [Ref. \onlinecite{kdotp}] & $0.54$               & $0.59$               & $0.35$               & $0.36$               \\
$A\,[{\rm \AA^2}]$ [Ref. \onlinecite{kdotp}] & $8.65$               & $9.37$               & $8.65$               & $9.37$               \\
$\frac{M_r}{m_p}$                         & $38.4$               & $59.7$               & $47.5$               & $85.0$                 \\
$\frac{M}{m_p}$                         & $160$               & $254$               & $248$               & $342$                               \\
$\hbar\omega_{LO}\,{\rm [meV]}$ [Ref. \onlinecite{defo_params}]           & $49$                 & $37$                 & $44$                 & $31$                 \\
$\hbar\omega_{HP}\,{\rm [meV]}$ [Ref. \onlinecite{defo_params}]           & $51$                 & $30$                 & $52$                 & $31$                 \\
$D_c\,{\rm [eV/ \AA]}$ [Ref. \onlinecite{defo_params}] & $5.8$                & $5.2$                & $3.1$                & $2.3$                \\
$D_v\,{\rm [eV/ \AA]}$ [Ref. \onlinecite{defo_params}] & $4.6$                & $4.9$                & $2.3$                & $3.1$                 \\
\hline \hline
\end{tabular}
\end{table}
To estimate the Born charge in Eq.\,(\ref{eq:glo}) we used DFT\cite{Baroni_2001} to calculate the Born effective charges of the atoms in the lattice of monolayer TMDCs.
The latter are defined by the response of the atomic displacements in a unit cell to a  homogeneous electric field. Hence, we write
\begin{equation} 
Z\equiv Z_{xx}=Z_{yy};\ Z_{ij}=\frac{1}{e}\frac{\partial F_j(s)}{\partial
  E_i}\biggr\rvert_{{\bf E}={\bf 0}},
\label{eq:born_again} 
\end{equation} where
${\bf F}(s)$ is the force acting on atom $s$ at its zero-field equilibrium
position. We used the \textsc{castep} plane-wave basis code
\cite{Clark_2005,Refson_2006} to calculate the Born effective charge tensors
for ${\rm MoS_2, MoSe_2, WS_2}$ and ${\rm WSe_2}$;  
\footnote{Note that Eq. (\ref{eq:born_again}) is
evaluated using Eqs.\ (40) and (42) of Ref.\ \onlinecite{Gonze_1997b}. To
evaluate Eq.\ (42) of Ref.\ \onlinecite{Gonze_1997b}, derivatives of the
Kohn--Sham orbitals with respect to the atomic positions and with respect to the wavevector are required.  The latter are evaluated within the
parallel-transport gauge by minimizing the functional in Eq.\ (70) of Ref.\ \onlinecite{Gonze_1997}.}
see Table \ref{table:born_charges}. We used the Perdew--Burke--Ernzerhof \cite{Perdew_1996} (PBE) exchange--correlation functional, norm-conserving pseudopotentials, a plane-wave cutoff energy of $\sim816\,{\rm eV}$, a $97 \times 97$ Monkhorst--Pack grid of ${\bf k}$-points, and (for the in-plane components of the Born effective charge tensors) an artificial (out-of-plane) periodicity of $\sim16\,{\rm \AA}$. We verified that our results are converged with respect to these parameters.
For the out-of-plane component we found a significant dependence on the artificial periodicity, which we removed by extrapolating to infinite layer separation.
\begin{figure}
\centering
\includegraphics[width=0.43\textwidth]{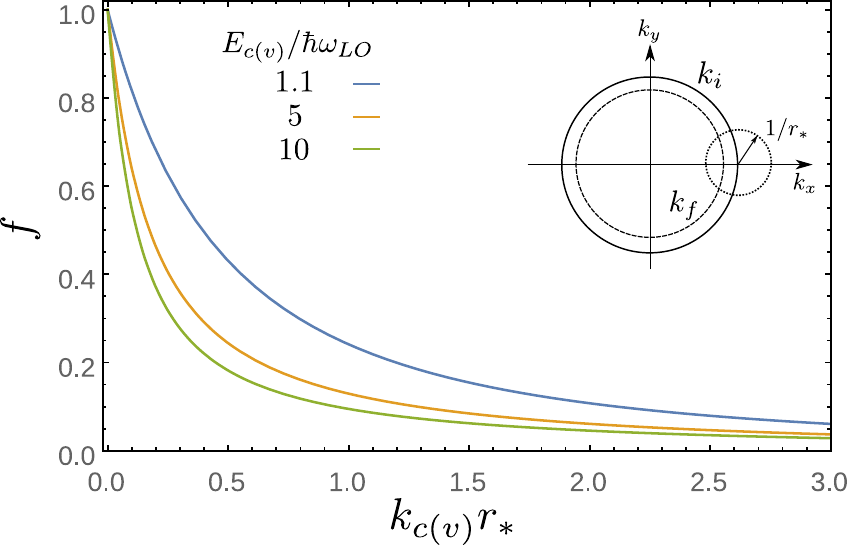}
\caption{The dimensionless function $f(k_{c(v)}r_*)$ in Eq.\,(\ref{eq:tau_lo}) for different carrier energies, $E_{c(v)}/\hbar\omega_{LO} = 1.1, 5$ and $10$. 
The inset sketches the kinematics for the phonon emission process in momentum space for the carrier undergoing energy relaxation with initial state wavevector $k_i$ and final-state wavevector $k_f$. The circle with radius $1/r_*$ defines the region of phonon wavevectors $q$ that give the dominant contribution to the scattering rate.  As the number of available final states scales as the circumference of the iso-energy circle in momentum space, for a given $r_*$ the asymptotic behavior of the scattering rate for high carrier energies is given by $\tau \sim 1/k_i \propto 1/\sqrt{E}$.}
\label{fig:scat_times}
\end{figure}
\begin{table}[th]
	\caption{Nonzero elements of the Born effective charge tensors $Z_{ij}$ of the transition-metal ions in TMDCs, as calculated using density-functional perturbation theory with the PBE functional, and the Born charge $Z$.
		\label{table:born_charges}}
	\vspace{0.2cm}
	\begin{tabularx}{0.8\columnwidth}{XYY}
		\hline \hline
		
		& $Z\equiv Z_{xx}=Z_{yy}$ & $Z_{zz}$ \\
		
		\hline
		
		MoS$_2$ & $-1.08$ & $-0.07$ \\
		
		MoSe$_2$ & $-1.80$ & $-0.09$ \\
		
		
		WS$_2$ & $-0.47$ & $-0.05$ \\
		
		WSe$_2$ & $-1.08$ & $-0.08$ \\
		
		\hline \hline
	\end{tabularx}
\end{table}

The homopolar (HP) mode (which corresponds to the irreducible representation $A_1'$ of the symmetry group $D_{3h}$) couples with the carriers through the lattice deformation potential\footnote{At the $\Gamma$-point, the TO and LO modes are degenerate, and the  displacement field of these two modes is the 2D vector ${\vec{u}}$ which transforms according to the representation $E'$ of the symmetry group $D_{3h}$. As a result, the pseudo-potential induced by the atomic displacements, which is a scalar function, should contain vector ${\vec{u}}$ in conjunction with another vector, which for the intra-valley scattering of electrons can only be a 2D gradient operator. Hence, the deformation potential of LO/TO mode would appear in the power-law expansion together with the wavevector ${\vec{q}}$ of the phonon, hence, it vanishes at $\Gamma$-point. For the LO phonon mode, the factor $q$, which appears through the gradient operator in the local displacement-induced charge density, $\vec{\nabla}\cdot{\vec{u}}$, is canceled by the $1/q$ factor coming from the 2D Fourier transform of the Coulomb potential, resulting in a finite contribution at the $\Gamma$-point in Eq.\,(\ref{eq:glo}).}
\begin{equation}
g_{HP}^{\alpha}=\sqrt{\frac{\hbar}{2NM\omega_{HP}}}D^{\alpha},\qquad \alpha = c\,\, \rm{or}\,\, v,
\end{equation}
where $M$ is the total atomic mass within the unit cell, $\omega_{HP}$ is the HP phonon frequency, and we distinguish electrons in the conduction band $(c)$ and holes in the valence band $(v)$. Here we follow the definitions given in Refs.\,\onlinecite{defo_params, defo_params2, phonon_mob} for the coupling and, below, we use the deformation potentials for the HP phonon mode reported in Ref. \onlinecite{defo_params}.
\begin{figure*}[ht]
	\includegraphics[width=0.8\textwidth]
	{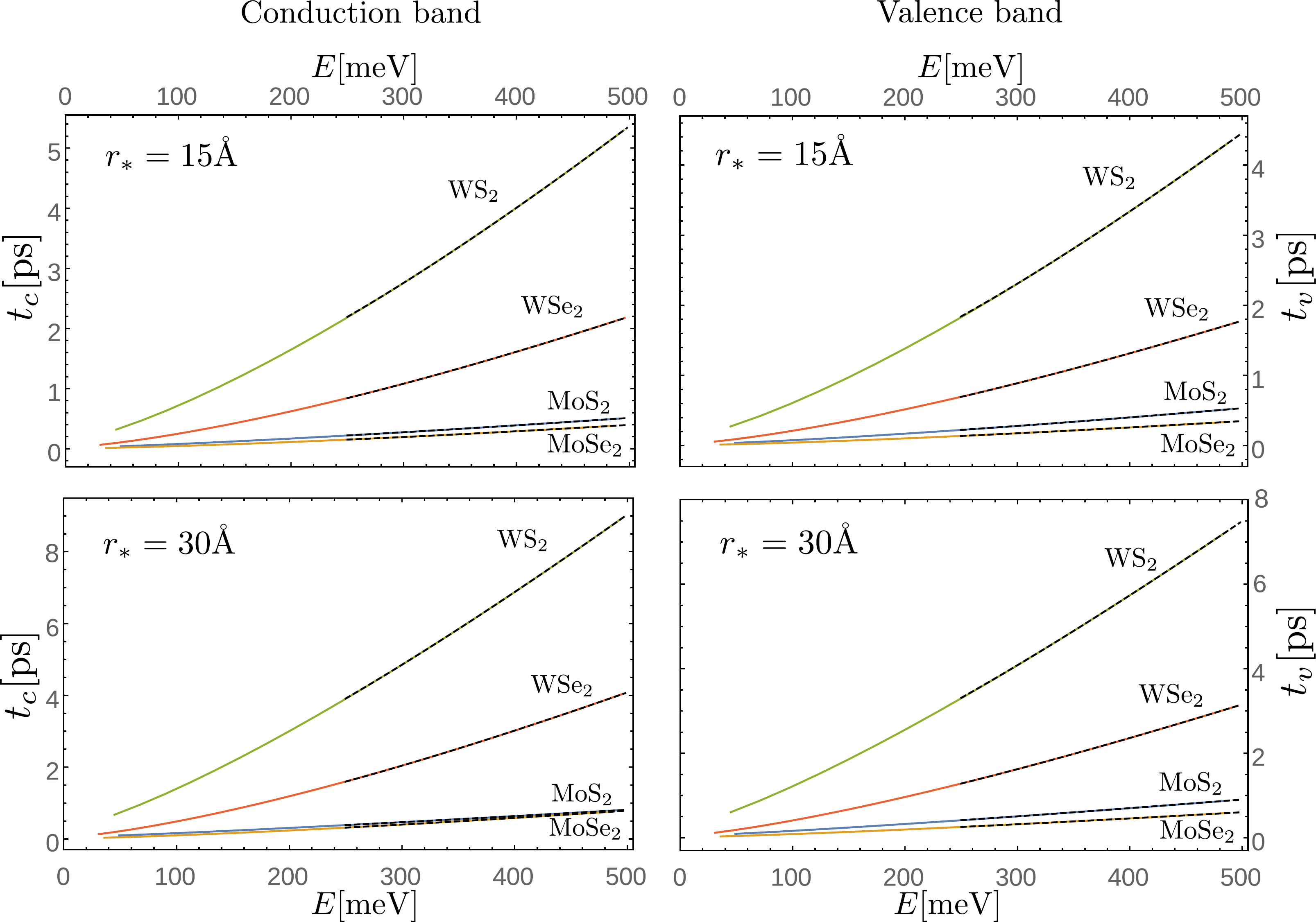}
	\caption{Hot carrier cooling time as a function of initial conduction-band (Left) and valence-band (Right) carrier energy $E$ for four TMDC materials, with $r_*=15{\rm \AA}$ (Top) and $r_*=30{\rm \AA}$ (Bottom). The dashed lines are asymptotic fits of the form in Eq.\,(\ref{eq:asym}).}
	\label{fig:rel_times}
\end{figure*}
The emission of both LO and HP phonons by a photo-excited electron/hole with initial momentum $k_i$ measured from the center of the corresponding ($K$ or $K'$) valley, is characterized by the rate calculated using the Fermi golden rule,
\begin{equation*}
\begin{split}
\label{eq:golden}
&\tau^{-1}=\frac{2\pi}{\hbar}\sum\limits_{\vec{q},\mu} |\langle f| H_{e-ph} |i \rangle|^2 \delta(E_f-E_i).
\end{split}
\end{equation*}
This yields
\begin{subequations}
\begin{equation}
\begin{split}
&\tau_{LO,\alpha}^{-1}=\tau_{\alpha}^{-1}
f\left(\frac{E_{\alpha}}{\hbar\omega_{LO}},k_{\alpha} r_*\right),\qquad \alpha = c\,\,\rm{or}\,\,v
\\
&\tau_{\alpha}^{-1}=\frac{2\pi^2 Z^2 E_B}{\hbar}\frac{m_{\alpha}}{M_r}\frac{a_B^2}{A}\frac{E_B}{\hbar\omega_{LO}};
\\
&f=\frac{1}{\pi}\frac{k_{\alpha}}{ k_i} \int\limits_{u_-}^{u_+}
\frac{du}{(1+uk_{\alpha}r_*)^2\sqrt{1-[\frac{k_{\alpha}}{2k_i}(u+\frac{1}{u})]^2}};
\\
&u_{\pm}=\frac{k_i}{k_{\alpha}}\left(1\pm \sqrt{1-\frac{{k_{\alpha}}^2}{k_i^2}}\right);
\quad k_{\alpha}=\sqrt{\frac{2m_{\alpha}\omega_{LO}}{\hbar}};
\end{split}
\label{eq:tau_lo}
\end{equation}
\begin{equation}
\tau^{-1}_{HP,\alpha}=\frac{m_{\alpha} A D^2_{\alpha}}{2M\hbar^2\omega_{HP}}.
\label{eq:thp}
\end{equation}
\end{subequations}
Note that these scattering rates are valid only for carrier energies above the corresponding optical phonon energy. 
Furthermore, the rate of emission of the HP phonon is independent of the carrier energy, due to the constant coupling coefficient and the constant density of states for 2D carriers with parabolic dispersion.
For the LO phonon mode we express the scattering rate in terms of a dimensionless integral by performing  a change of variables, defining the dimensionless variable $u=q/k_{c(v)}$, where $k_{c(v)}$ is the carrier wavevector corresponding to an energy of $\hbar\omega_{LO}$, and $a_B$ and $E_B$ are the Bohr radius and energy. In Table\,\ref{params} we list the values of the parameter $\tau^{-1}_{c(v)}$ for various TMDCs and in  Fig.\,\ref{fig:scat_times} we show the shape of the function $f$ for different carrier energies $E_{c(v)}$. The decrease of this scattering rate upon increasing $r_*$ or excitation energy can be understood from the diagram depicting the kinematic phase space for a carrier emitting an optical phonon. 

Comparing the values of $\tau_{c(v)}^{-1}$ and $\tau_{HP,c(v)}^{-1}$ in Eqs.\,(\ref{eq:tau_lo}), (\ref{eq:thp}), and Table\,\ref{params_fit}, we see that emission of the LO phonon mode dominates in the relaxation over the HP phonon. The two rates become comparable for sufficiently large carrier energies or sufficiently large $r_*$ values. Asymptotically, we have for the LO phonon, $\tau_{LO}^{-1}\sim 1/(r_*\sqrt{E})$; therefore, the boundary between the two modes is given by\footnote{This parameter was derived by equating $\tau_{LO}^{-1}\sim \tau_{HP}^{-1}$, and using the asymptotic form of $\tau_{LO}^{-1}$ for large carrier energies. The corresponding values are, $16, 38, 15$ and $82$ for electrons and $26, 43, 28$ and $45$ for holes in ${\rm MoS_2, MoSe_2, WS_2,}$ and ${\rm WSe_2}$, respectively.} $\frac{r_*\sqrt{m E}}{\hbar} \sim 4\pi^2 Z^2 \frac{a_B^2E_B^2}{A^2D^2}\frac{M}{M_r}\frac{\omega_{HP}}{\omega_{LO}}$, where $E$ is the energy of the photo-excited carrier and $m$ is its band mass.
%

For hot carriers excited to the energy $E\gg \hbar\omega_{LO/HP}$, we write the cooling rate as
\begin{subequations}
\label{eq:trel}
\begin{equation}
\frac{dE}{dt}=-\frac{\hbar\omega_{LO}}{\tau_{LO}(E)}
-\frac{\hbar\omega_{HP}}{\tau_{HP}},
\end{equation}
so that we can determine the relaxation time as a function of the initial carrier energy $E$ as\footnote{For $\epsilon<\hbar\omega_{LO}$, we set $\tau_{HP}^{-1}=0$, and integrate the extrapolated scattering rate of the LO phonon mode for $\epsilon<\hbar\omega_{LO}$ to obtain the total relaxation time.}
\begin{equation}
t(E\gg\hbar\omega_{LO})=
\int\limits_{0}^{E} \frac{d\epsilon}{\frac{\hbar\omega_{LO}}{\tau_{LO}(\epsilon)}+\frac{\hbar\omega_{HP}}{\tau_{HP}}}.
\end{equation}
For hot carriers excited to the energy $E\gg \hbar\omega_{LO/HP}$, Eq.\,(\ref{eq:tau_lo}) yields $\tau^{-1}_{LO}\propto 1/\sqrt{E}$ and shows that $\tau_{HP}^{-1}$ is constant (also see Fig.\,\ref{fig:scat_times}), so that we find an analytical asymptotic form for the cooling time of charge carriers from the initial energy $E$ to the bottom of the band,
\begin{equation}
t(E)\approx aE-b\sqrt{E}+c,
\label{eq:asym}
\end{equation}
\end{subequations}
where the evaluated values of the parameters $a, b$ and $c$ are listed in Table\,\ref{params_fit} and correspond to the numerically obtained\cite{zhang} relaxation time curves shown in Fig.\,\ref{fig:rel_times} for the conduction and valence bands with $r_*= 15{\rm \AA}$ and $30{\rm \AA}$. 

\begin{table}[ht]
\centering
\caption{Values of relevant parameters in Eqs.\,$(\ref{eq:tau_lo}),(\ref{eq:thp})$, and $(\ref{eq:asym}$). Listed are the carrier wavevectors $k_{c(v)}$, for the conduction (c) and valence (v) bands corresponding to a carrier energy of $\hbar\omega_{LO}$, the scattering rate $\tau_{c(v)}^{-1}$ due to LO phonon emission with $r_*=0$, and the constant scattering rates $\tau_{HP,c(v)}^{-1}$ due to HP phonon emission.
The bottom parts list the fitting parameters $a_{c(v)}, b_{c(v)}$ and $c_{c(v)}$ for the asymptotic forms of the relaxation times as a function of the carrier energy for $r_*=15 {\rm \AA}$ and $r_*=30 {\rm \AA}$ for the conduction (c) and valence (v) bands.}
\vspace{0.2cm}
\label{params_fit}
\begin{tabular}{lccccc}
\hline\hline                                          & ${\rm MoS_2}$      & ${\rm MoSe_2}$     & ${\rm WS_2}$ & ${\rm WSe_2}$      & \multicolumn{1}{l}{}                \\ \hline
$k_c\,{\rm [\AA^{-1}]}$                 & $0.077$            & $0.074$            & $0.055$      & $0.048$            & \multicolumn{1}{l}{}                \\
$k_v\,{\rm [\AA^{-1}]}$                 & $0.083$            & $0.076$            & $0.064$      & $0.054$            & \multicolumn{1}{l}{}                \\
$\tau^{-1}_c\,{\rm [ps^{-1}]}$            & $112$              & $296$              & $11$         & $45$               & \multicolumn{1}{l}{}                \\
$\tau^{-1}_v {\rm [ps^{-1}]}$             & $130$              & $312$              & $14$         & $58$               & \multicolumn{1}{l}{}                \\
${\tau_{HP,c}}^{-1} {\rm [ps^{-1}]}$    & $6.8$              & $7.7$              & $0.69$       & $0.54$             & \multicolumn{1}{l}{}                \\
${\tau_{HP,v}}^{-1} {\rm [ps^{-1}]}$    & $5.0$              & $7.2$              & $0.5$        & $1.3$              & \multicolumn{1}{l}{}                \\
\hline\hline
$a_c\,{\rm [\frac{ps}{meV}]}$        & $1.8\cdot10^{-3}$  & $1.8\cdot 10^{-3}$ & $0.021$      & $9.5\cdot10^{-3}$  & \multirow{6}{*}{$r_*=15 {\rm \AA}$} \\
$b_c\,{\rm [\frac{ps}{\sqrt{meV}}]}$ & $0.025$            & $0.031$            & $0.34$       & $0.16$             &                                     \\
$c_c\,{\rm [ps]}$                    & $0.16$             & $0.19$             & $2.1$        & $0.93$             &                                     \\
$a_v\,{\rm [\frac{ps}{meV}]}$        & $2.0\cdot10^{-3}$  & $1.5\cdot10^{-3}$  & $0.017$      & $7.4\cdot 10^{-3}$ &                                     \\
$b_v\,{\rm [\frac{ps}{\sqrt{meV}}]}$ & $0.031$            & $0.024$            & $0.27$       & $0.12$             &                                     \\
$c_v\,{\rm [ps]}$                    & $0.20$             & $0.14$             & $1.67$       & $0.68$             &                                     \\ \hline
$a_c\,{\rm [\frac{ps}{meV}]}$        & $2.3\cdot 10^{-3}$ & $3.2\cdot 10^{-3}$ & $0.031$      & $0.017$            & \multirow{6}{*}{$r_*=30 {\rm \AA}$} \\
$b_c\,{\rm [\frac{ps}{\sqrt{meV}}]}$ & $0.023$            & $0.051$            & $0.42$       & $0.26$             &                                     \\
$c_c\,{\rm [ps]}$                    & $0.17$             & $0.32$             & $2.6$        & $1.5$              &                                     \\
$a_v\,{\rm [\frac{ps}{meV}]}$        & $2.8\cdot10^{-3}$  & $2.3\cdot10^{-3}$  & $0.025$      & $0.012$            &                                     \\
$b_v\,{\rm [\frac{ps}{\sqrt{meV}}]}$ & $0.032$            & $0.032$            & $0.31$       & $0.17$             &                                     \\
$c_v\,{\rm [ps]}$                    & $0.22$             & $0.20$             & $1.96$       & $0.96$             &                                     \\ \hline\hline
\end{tabular}
\end{table}

In summary, we calculated the scattering rates and relaxation times of photo-excited carriers in TMDCs due to optical phonon emission. We obtained relaxation times of a few ${\rm ps}$ for all the materials studied, with ${\rm MoSe_2}$ having the shortest relaxation time for all carrier energies up to $0.5\,{\rm eV}$. These results demonstrate that TMDCs feature short time-scale energy relaxation.
\begin{acknowledgements}
We thank M. Calandra, T. Heinz,  K. Novoselov, M. Potemski, A. Tartakovskii, and V. Z\'{o}lyomi for useful discussions. This work was supported by the Simons Foundation, the ERC Synergy Grant Hetero2D and the EC-FET European Graphene Flagship.
\end{acknowledgements}
\bibliographystyle{apsrev4-1} 
\bibliography{bibp}
\end{document}